\documentclass[lettersize,journal]{IEEEtran}
\usepackage{amsmath,amsfonts}
\usepackage{algorithmic}
\usepackage{algorithm}
\usepackage{array}
\usepackage[caption=false,font=normalsize,labelfont=sf,textfont=sf]{subfig}
\usepackage{textcomp}
\usepackage{stfloats}
\usepackage{url}
\usepackage{verbatim}
\usepackage{graphicx}
\usepackage{cite}
\usepackage{CJKutf8}
\usepackage{enumitem}
\usepackage{color}
\usepackage{boxedminipage}
\usepackage{fancybox}
\usepackage{multirow}
\usepackage{amsmath}
\usepackage{amsfonts}
\usepackage{amsthm}
\usepackage{mathtools}
\usepackage{diagbox}
\usepackage{tabularx}
\usepackage{pifont}
\usepackage{threeparttable}
\usepackage{subfig}
\usepackage{lastpage}
\usepackage{pdflscape}
\usepackage{longtable}
\usepackage{listings}

\theoremstyle{definition}

\newcolumntype{P}[1]{>{\centering\arraybackslash}p{#1}}

\DeclarePairedDelimiter\abs{\lvert}{\rvert}%
\DeclarePairedDelimiter\norm{\lVert}{\rVert}%
\makeatletter
\let\oldabs\abs
\def\abs{\@ifstar{\oldabs}{\oldabs*}}
\let\oldnorm\norm
\def\norm{\@ifstar{\oldnorm}{\oldnorm*}}
\makeatother

\begin{document}

\title{Qualcomm Trusted Application Emulation for Fuzzing Testing}

\author{Chun-I Fan,~\IEEEmembership{Senior Member,~IEEE,}
  Li-En Chang,
  Cheng-Han Shie\IEEEauthorrefmark{1},
\IEEEcompsocitemizethanks{\IEEEcompsocthanksitem  Chun-I Fan is with the Department of Computer Science and Engineering, National Sun Yat-sen University, Kaohsiung 804201, Taiwan (E-mail: cifan@mail.cse.nsysu.edu.tw).
\IEEEcompsocthanksitem Li-En Chang is with the Department of Computer Science and Engineering, National Sun Yat-sen University, Kaohsiung 804201, Taiwan (E-mail: yiblueli@gmail.com).
\IEEEcompsocthanksitem Cheng-Han Shie is with the Department of Computer Science and Engineering, National Sun Yat-sen University, Kaohsiung 804201, Taiwan (E-mail: hanhan3927@g-mail.nsysu.edu.tw).
\\
}

\thanks{This work was supported in part by the National Science and Technology Council of Taiwan under Grants NSTC 113-2634-F-110-001-MBK and 113-2221-E-110-082, in part by the Information Security Research Center at National Sun Yat-sen University, and in part by the Intelligent Electronic Commerce Research Center from the Featured Areas Research Center Program through the Framework of the Higher Education Sprout Project by the Ministry of Education in Taiwan.

\IEEEauthorrefmark{1}Corresponding author.}

}



\maketitle
\begin{abstract}
In recent years, the increasing awareness of cybersecurity has led to a heightened focus on information security within hardware devices and products. Incorporating Trusted Execution Environments (TEEs) into product designs has become a standard practice for safeguarding sensitive user information. However, vulnerabilities within these components present significant risks, if exploited by attackers, these vulnerabilities could lead to the leakage of sensitive data, thereby compromising user privacy and security. This research centers on trusted applications (TAs) within the Qualcomm TEE and introduces a novel emulator specifically designed for these applications. Through reverse engineering techniques, we thoroughly analyze Qualcomm TAs and develop a partial emulation environment that accurately emulates their behavior. Additionally, we integrate fuzzing testing techniques into the emulator to systematically uncover potential vulnerabilities within Qualcomm TAs, demonstrating its practical effectiveness in identifying real-world security flaws. This research makes a significant contribution by being the first to provide both the implementation methods and source codes for a Qualcomm TAs emulator, offering a valuable reference for future research efforts. Unlike previous approaches that relied on complex and resource-intensive full-system simulations, our approach is lightweight and effective, making security testing of TA more convenient.

\end{abstract}

\section{Introduction}

A Trusted Execution Environment (TEE) is an isolated execution environment designed to enhance system security. It is widely utilized in mobile devices, smart cards, IoT devices, and various other applications. Originally proposed by ARM, TEE is implemented in the ARM TrustZone security architecture, which is prevalent in ARM architecture processors. TEE leverages system-wide hardware to create a secure execution environment, safeguarding the confidentiality and integrity of the code and data within it. As security needs and technology advance, the scope of TEE has expanded to encompass smartphones, smart cards, wearable devices, and IoT devices. Traditional operating systems may suffer from system crashes or security risks due to application vulnerabilities. 

In contrast, TEE provides a protected environment that ensures the overall system security remains intact, even if an application contains vulnerabilities. Additionally, TEE secures digital content by protecting copyright and encryption keys, thus preventing unauthorized access and copying—a crucial feature for businesses offering digital content services. TEE effectively addresses the security requirements of various devices. Despite its benefits, security vulnerabilities in TEE can be exploited by attackers through memory corruption or other memory-related issues, such as buffer overflows or format string attacks. While TEE aims to protect sensitive user information, vulnerabilities can lead to information leaks, exposing devices to high-risk environments.

As the use of TEE continues to grow, its internal security has become a critical area of research. Recent studies have highlighted vulnerabilities within the TEE environment. The following section explores current research on TEE, focusing on fuzzing methods applied to both physical devices and virtual environments.

To enhance the reproducibility of our research and foster further advancements in TEE security, we have made the implementation of our Qualcomm TAs emulator and the associated fuzzing techniques publicly available at \cite{theproposedemu}.

\section{Preliminaries}
\subsection{ARM TrustZone}
\label{sec:TZ}
\begin{figure}[tbp]
    \centering
    \includegraphics[width = \columnwidth]{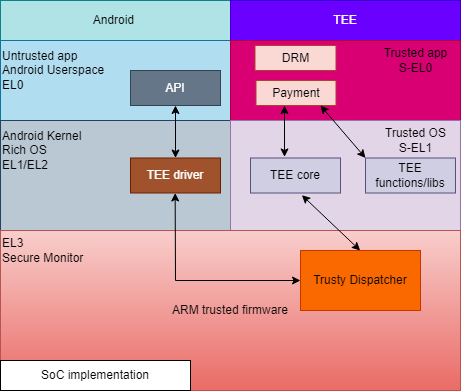}
    \caption{ARM TrustZone}
    \label{fig:tz_re}
\end{figure}

TrustZone ( Figure\ref{fig:tz_re} ), developed by ARM, is a comprehensive security solution that operates at both the software and hardware levels. It provides a hardware-based protection mechanism to secure sensitive data and system functions on a device. TrustZone uses hardware separation technology to create two independent execution environments on the same processor: the secure execution environment (secure world) and the normal execution environment (normal world). These environments are isolated, each with its processor state, memory space, and system resources, ensuring that sensitive information and operations remain protected from the normal environment.

TrustZone defines permissions across several levels, ranging from Exception Level 0 (EL0), the lowest, to Exception Level 3 (EL3), the highest. EL0 is user mode, accessed by general users. EL1 is RichOS in the normal world, handling system calls. EL2 is hypervisor mode, seldom used in typical mobile devices. EL3, secure monitor mode, holds the highest permission level, managing access between the normal world and the secure world. The secure world operates at Secure Exception Level 0 (S-EL0) and S-EL1.

\subsection{An Overview of Emulation Technologies}
\label{sec:Emulation_overview}
QEMU, Unicorn, and Qiling are widely utilized virtualization and emulation tools in software development, security research, and testing. These tools provide virtual hardware environments, enabling developers to test software across different platforms and architectures, which significantly reduces hardware costs and development time. Security researchers use these tools to conduct vulnerability exploitation experiments, uncover new vulnerabilities, and identify attack vectors, leading to the development of effective defensive measures.

\noindent\textbf{Qemu} \cite{Qemu} is an open-source virtualization tool supporting the emulation of multiple hardware architectures, including x86, ARM, and MIPS. It emulates a complete system, encompassing the processor, memory, and peripheral devices, and runs multiple virtual machines of different architectures. QEMU employs binary translation to emulate various hardware architectures and supports hardware acceleration, like KVM, to enhance performance.

\noindent\textbf{Unicorn} \cite{unicorn} is a lightweight, open-source instruction set emulator used in binary analysis, vulnerability exploitation, and related fields. Unicorn emulates and executes individual instructions, functions, applications, and operating systems, with support for dynamic symbol resolution, memory mapping, and analysis. It uses dynamic binary translation to emulate instruction sets across architectures and handles system calls and external devices through virtual memory and callback functions.

\noindent\textbf{Qiling} \cite{Qiling-Framework} is an open-source emulation framework designed to streamline and accelerate tasks like binary analysis, vulnerability research, and security testing. Built on the Unicorn emulation engine, Qiling offers higher-level APIs and tools, simplifying development and testing. It supports multiple instruction set architectures, such as x86, ARM, and MIPS, and provides Python and C APIs for easy integration into existing tools.

\subsection{Fuzzing Testing}
\label{sec:Fuzzing Testing}

Fuzzing testing and unit testing are two software testing methods. Fuzzing seeks to uncover unexpected behaviors or vulnerabilities by injecting numerous random or abnormal inputs into the program. This approach identifies input conditions that trigger abnormal behavior, crashes, or security vulnerabilities, focusing primarily on security-related issues.

Unit testing, on the other hand, verifies the correctness of individual units (such as functions, methods, or classes) in the software. It requires a deep understanding of the program design and tests specific functions based on defined inputs and expected results. Unit testing is conducted early in the development process and is often integrated into the continuous integration and deployment (CI/CD) pipeline to ensure code correctness.

Fuzzing testing is generally divided into three categories:

\begin{itemize}
    \item \textbf{Black-box fuzzing:} Black-box fuzzing is an automated security testing technique designed to uncover vulnerabilities and security weaknesses in software applications. Unlike white-box fuzzing, black-box fuzzing focuses solely on input and output, disregarding the internal code structure. The seed mutation strategy involves mutating the initial seed input to generate diverse test cases, ensuring that the generated cases effectively cover various scenarios and input ranges of the application.
    \item \textbf{White-box fuzzing:} White-box fuzzing allows testers to have a deeper understanding of the software's internal implementation. The fuzzer conducts targeted testing based on the software's structure and code, such as testing specific functions or code paths.
    \item  \textbf{Gray-box fuzzing:} Gray-box fuzzing strikes a balance between black-box and white-box fuzzing. In gray-box fuzzing, the tester possesses limited internal knowledge of the target software or system, typically including code structure, function calls, and key variables. This information may be acquired through static analysis, reverse engineering, or other methods. Overall, gray-box fuzzing combines the universality and flexibility of black-box testing with the precision and effectiveness of white-box testing, making it an effective approach for discovering software vulnerabilities and errors.
\end{itemize}

Seed Mutation: During fuzzing, new test cases are generated by modifying, mutating, or transforming the initial test cases (seeds). These new cases retain the basic characteristics and structure of the original seeds but may vary slightly to explore different execution paths.
\begin{itemize}
    \item \textbf{Random mutation:} The most basic mutation strategy, where seed input is randomly altered to produce new test cases. This can involve inserting, deleting, or replacing specific bytes, strings, or other input elements.
    \item \textbf{Dictionary-based mutation:} This strategy utilizes a predefined dictionary or vocabulary to generate mutated test cases. The dictionary typically includes common strings, file formats, or protocol messages. New test cases are created by randomly selecting and combining elements from the dictionary.
    \item \textbf{Pattern matching-based mutation:} Focuses on identifying and exploiting patterns or structures within the input. By analyzing known input formats, the strategy generates mutated test cases based on these patterns, aiming to uncover vulnerabilities or edge cases.
    \item \textbf{Mutation probability-based mutation:} Adjusts mutations based on the probability assigned to each element. Elements with higher mutation probability are more frequently altered, while those with lower probability might be modified minimally or not at all.
    \item \textbf{Structure-aware mutation:} Based on an understanding of the application’s structure, this strategy mutates different structural elements. For instance, when working with a composite data structure, the focus might be on altering individual members rather than the entire structure.
\end{itemize}

\subsection{Unicorn Engine}
\label{sec:Unicorn}

    As the number of IoT (Internet of Things) devices grows, the need for IoT security research increases. However, the processor instruction sets used in IoT devices often differ from those on our local computers, presenting research challenges. For instance, in the Android system, most programs run on AArch64 or ARM architecture instruction sets. When conducting tests and research, if the target program's source code is available, it can be compiled into a program with the same local instruction set architecture for testing. However, many programs are closed source, making it crucial to execute programs of different architectures without access to the source code. Common emulation methods include using tools like QEMU, Unicorn, and Qiling. This research focuses on the Unicorn engine for its emulation strategy.\\

    The Unicorn engine is an open-source command emulator framework designed specifically to execute binary code across various architectures. This engine provides a flexible platform that allows developers to quickly and easily build and execute instruction-level emulators while supporting a wide range of processor architectures, including x86, ARM, MIPS, SPARC, and PowerPC. Unicorn Engine can operate on multiple platforms, such as Linux, Windows, and macOS. Its ability to perform instruction-level emulation in different environments offers an adaptable framework that lets developers extend and customize the engine's functionality with ease. It supports user-defined instruction set extensions and includes various plugins and modules, such as input/output emulation, memory emulation, and system call emulation.\\

    Instruction set conversion is achieved using intermediate representation (IR) and tiny code generation (TCG). TCG first parses and decodes the virtual instruction set, generating equivalent native machine code in basic block units. For example, the CPU instruction set for general computers is x86-64. To execute the AArch64 instruction set on x86-64, TCG conversion is required to translate AArch64 instructions into those executable by x86-64. Additionally, the Unicorn engine offers powerful hook technology, enabling users to implement numerous hooks through APIs to maintain the emulation architecture's integrity and dynamically debug emulation environment issues.

\subsection{Vendor of TEE }
\label{sec:Vendor of TEE}

\noindent\textbf{ARM (Advanced RISC Machines)} is a leading semiconductor design company that provides TrustZone technology \cite{Arm-TrustZone}, one of the most widely used solutions in TEE technology. ARM's TrustZone technology is extensively used in embedded systems, including smartphones and IoT devices.

\noindent\textbf{GlobalPlatform} \cite{GlobalPlatform} is an international non-profit organization focused on developing and promoting secure smart card technology standards, including TEE technology. GlobalPlatform's specifications offer a common standard for various vendors to develop and deploy compliant TEE solutions.

\noindent\textbf{Trustonic} \cite{Trustonic} is a security technology company offering TEE solutions based on ARM TrustZone technology. Trustonic's products are widely used in smartphones, payment terminals, and other devices to protect sensitive information and provide a secure environment for application execution. Kinibi and TEEGRIS are TEEOS developed by Trustonic, mainly used in Samsung mobile phones with Exynos chips. Kinibi was predominantly used in mobile phones before the Samsung S10 series, while TEEGRIS is the trusted execution environment used in devices after the S10 series.

\noindent\textbf{Qualcomm} \cite{Qualcomm} is a renowned semiconductor company that offers security solutions based on ARM TrustZone technology, integrated into the Snapdragon series of processors and chipsets. Qualcomm TEE supports a secure boot process and maintains a complete chain of trust, ensuring that software and code in the TEE remain unaltered and untampered when loading and executing.

\section{Related Works}
\subsection{TEEzz: Fuzzing TAs on COTS Android Devices}
\label{sub:TEEzz}

Busch et al. \cite{10179302} performed fuzzing tests on TAs within TEE using physical Android devices. They introduced a method to identify and exploit vulnerabilities by fuzzing the TEE interface, including system calls and drivers, and utilizing TEE’s error-handling mechanisms. However, testing on physical devices creates a black-box scenario, which limits the analysis of internal memory. To overcome this, they used Client Applications (CA) to understand the expected data types at the interface and generate type-aware seeds for fuzzing.

This method effectively identifies vulnerabilities in TEE on commercial off-the-shelf (COTS) Android devices. Type-aware seeds improve fuzzing efficiency and accuracy, potentially uncovering subtle and complex vulnerabilities.

The black-box nature of physical devices restricts detailed internal memory analysis, which may overlook vulnerabilities only visible with internal state access. Additionally, using physical devices can make the testing process more cumbersome and time-consuming compared to virtualized environments.

\subsection{PartEmu: Enabling Dynamic Analysis of Real-World TrustZone Software Using Emulation}

\label{sub:PartEmu}

Harrison et al. \cite{247658} introduced PARTEMU, a tool designed for the dynamic analysis of software within the ARM TrustZone environment. They observed that most ARM TrustZone research emphasizes static analysis and emulation, lacking effective tools for dynamic analysis of actual software. PARTEMU addresses this gap by emulating the TrustZone environment, allowing for real-time dynamic analysis during software execution and offering an extensible analysis framework. However, it is a complex and large-scale emulation environment, primarily supporting Trustonic TrustZone by emulating the bootloader, Secure Monitor, TEE Driver, and TEEOS.

PARTEMU provides a robust platform for dynamic analysis in a controlled environment, offering valuable insights into the runtime behavior of TrustZone software. Its extensible framework facilitates adaptation and expansion for various analysis needs, making it a versatile tool for in-depth security research.

The complexity and scale of the PARTEMU environment may demand substantial resources and expertise for setup and maintenance. Moreover, its focus on Trustonic TrustZone might limit its applicability to other TrustZone implementations, potentially affecting its generalizability.

\subsection{SoK: Understanding the Prevailing Security Vulnerabilities in TrustZone-assisted TEE Systems.}
\label{sub:Sok}

Cerdeira et al. \cite{9152801} investigated and analyzed the main security vulnerabilities in TrustZone-assisted TEE systems. They evaluated the security and trustworthiness of existing TEE systems through a comprehensive analysis of published vulnerability reports and related research. The study provides an overview of common attack vectors, vulnerability types, impact scope, and the effectiveness of existing defense mechanisms. Its main contribution is a comprehensive assessment of these security vulnerabilities and the proposal of a unified analysis framework to aid researchers in understanding and evaluating TEE system security. Offering a thorough and systematic analysis of security vulnerabilities in TrustZone-assisted TEE systems, this research serves as a valuable resource for both researchers and practitioners. The proposed unified analysis framework introduces a standardized approach for assessing TEE security, which can facilitate future research and enhance the design of more secure TEE systems. However, the focus on existing vulnerability reports and published research may limit coverage to known vulnerabilities, potentially overlooking emerging threats or novel attack vectors. Additionally, while the unified analysis framework is beneficial, its effectiveness in real-world scenarios may require further validation and refinement.

\subsection{DTA: Run TrustZone TAs Outside the secure world for Security Testing}
\label{sub:Sok}

Juhyun et al. \cite{song2024dta} proposed an approach called DTA to facilitate fuzzing testing of trusted applications by executing them outside the secure world. Traditional security testing methodologies for TAs face limitations due to the strict isolation and control mechanisms enforced by the secure world. DTA leverages dynamic binary translation and emulation techniques to relocate TAs into the normal world, enabling comprehensive security analysis and testing without compromising the secure world's integrity.

DTA operates by extracting trusted applications from the secure world and executing them in the normal world. This involves using dynamic binary translation to convert the TA's binary code for execution in the normal world and emulation techniques to mimic the secure world's environment. This relocation allows testers to apply standard security testing tools and methodologies without breaching the secure world’s security guarantees.

\subsection{Finding 1-Day Vulnerabilities in trusted applications using Selective Symbolic Execution}
\label{sub:1-day}

Busch et al. \cite{busch2020finding} introduced SimTA, a method for discovering 1-day vulnerabilities in real-world trusted applications using selective symbolic execution. SimTA, based on the angr emulator, emulates the execution environment of TAs. By leveraging insights from manual static analysis and exploiting a physical device, they demonstrated SimTA's effectiveness through binary-diff-guided analysis, successfully reproducing a critical vulnerability.

SimTA operates by selectively applying symbolic execution to the most relevant code sections, thereby reducing the complexity and resource demands typically associated with full symbolic execution. The approach uses binary-diff-guided analysis to concentrate on code regions altered between versions, making it especially effective for identifying 1-day vulnerabilities. Integrating SimTA with an emulated environment based on angr allows for accurate analysis of the TA execution context without requiring actual hardware.

Building on the insights gained from previous studies, this research aims to address some of the limitations identified in existing approaches, particularly in the context of dynamic analysis and emulation of trusted applications. While prior works such as PARTEMU \cite{247658} and DTA \cite{song2024dta} have made significant strides in enabling dynamic analysis and security testing of TAs, there remains a need for a more accessible, open-source solution that simplifies the emulation process without sacrificing accuracy or comprehensiveness. 

\section{The Proposed Scheme}
\label{ch:Proposed Scheme}

\begin{figure}[tbp]
  \centering
  \includegraphics[width=0.8\columnwidth]{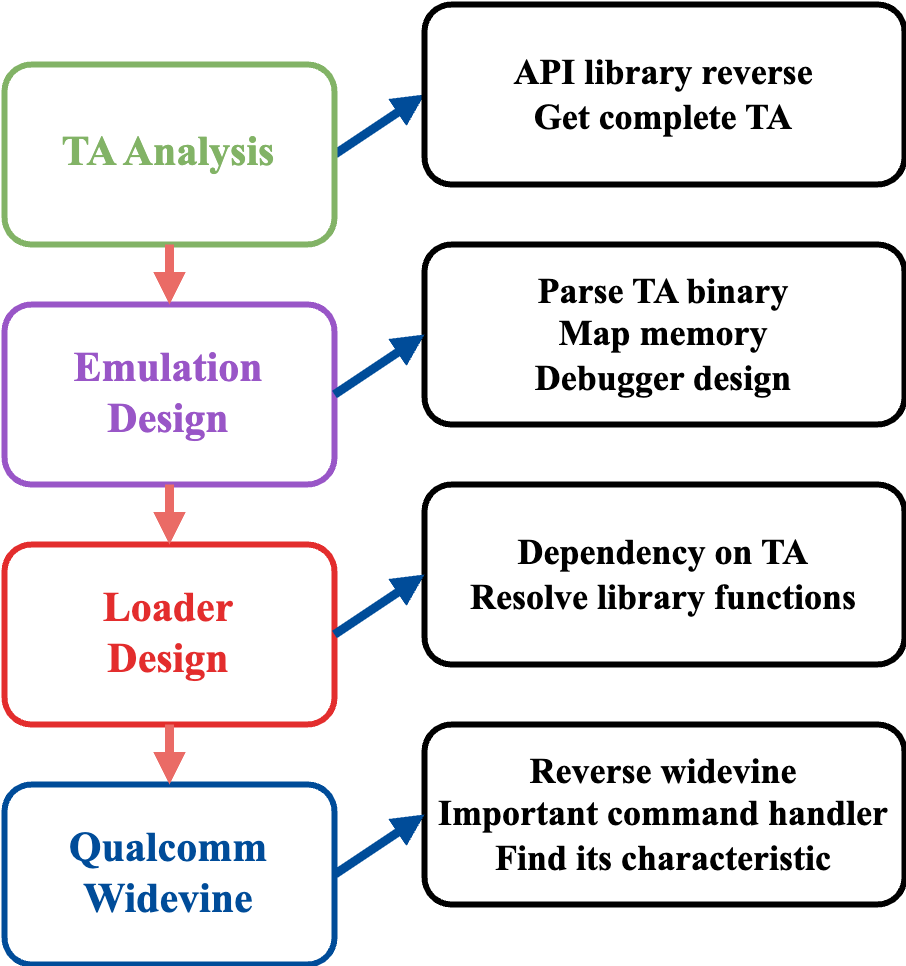}
  \caption{Scheme}
  \label{fig:total_flowchart}
\end{figure}

This chapter introduces the proposed scheme shown in Figure \ref{fig:total_flowchart} and details the process of extracting the trusted application and merging scattered files into a complete file. It covers various reverse engineering techniques. Additionally, the design of the emulator will be discussed, including the creation of a loader to map the contents of the shared object file. The chapter will also present the main focus of widevine trusted application, along with its internal reverse analysis. This research introduces a lightweight emulator designed for emulating and fuzzing TA functions. Unlike Harrison et al. \cite{247658}, which involves emulating numerous components and relies on closed-source methods that are difficult to replicate, this scheme offers an open-source emulation approach. This enables researchers to efficiently test TA functions for potential vulnerabilities. It is important to note that some code presented in the following \texttt{listing} format is decompiled through reverse engineering techniques, not open-source code.

\subsection{Trusted Application Analysis}

This research focuses on Qualcomm TAs, given the numerous weaknesses and vulnerabilities highlighted in recent studies \cite{9152801}. TAs are crucial to the TEE environment, implementing functions to safeguard user privacy. In 2015, related studies \cite{Explore_Qualcomm} conducted a reverse analysis of Qualcomm TAs. However, the architecture of Qualcomm TAs has since evolved, differing from earlier research. Consequently, a reverse analysis was performed to understand the current state of TAs.

\begin{figure}[h!]
  \centering
  \includegraphics[width=\columnwidth]{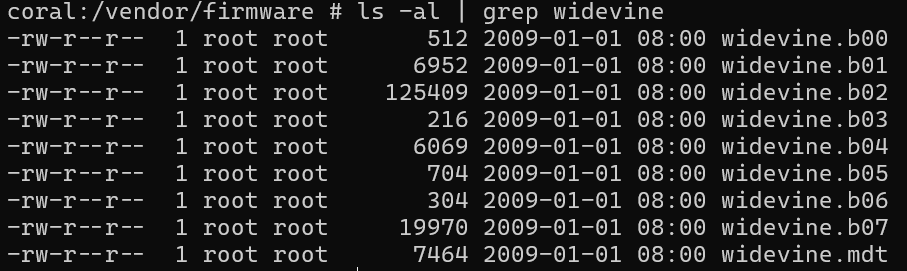}
  \caption{Widevine Files}
  \label{fig:widevine_files}
\end{figure}

First, locate the trusted application in the file system of the physical phone. As shown in Figure \ref{fig:widevine_files}, the directory contains many scattered files rather than a single executable file.

Understanding how a trusted application is loaded from the Normal World to the Secure World is crucial. Research reveals that the kernel module in the Rich OS is used to load the trusted application into the Secure World. This process utilizes the API in the shared library \texttt{libQSEEComAPI.so}.

Reverse analysis of \texttt{libQSEEComAPI.so} indicates that the \texttt{/dev/qseecom} driver facilitates communication between the normal world and the secure world. The API entry point is \texttt{QSEECom\_start\_app\_V2()}, responsible for reading the scattered trusted application files. The function \texttt{\_QSEECom\_load\_image()} reads the header of \texttt{TA\_NAME.mdt} and parses the program header to determine if the file is 32-bit or 64-bit. It also identifies the number of segments, which should match the number of scattered \texttt{TA\_NAME.b0x} files as shown in Listing \ref{load_ta}. The logic involves placing the contents of each file into the corresponding memory locations and merging them into a single executable file.

\begin{lstlisting}[language=C, caption=\texttt{\_QSEECom\_load\_image()}, label = load_ta ]
[...]
        arch = (*fread_buf)[4];
        if ( arch == 2 )
        {
          bxx_num = *&(*fread_buf)[0x38];       // 64bit -> 0x38
        }
        else
        {
          if ( arch != 1 )
          {
            __android_log_print(6LL, "QSEECOMAPI", "Error:: Invalid tz app architecture %d", (*fread_buf)[4]);
            goto LABEL_12;
          }
          bxx_num = *&(*fread_buf)[0x2C];       // 32bit -> 0x2c
        }
        if ( bxx_num == 0xFFFF )
        {
          __android_log_print(6LL, "QSEECOMAPI", "file %s 's total seg num %x is too big", &buf.file[1], 0xFFFFLL);
          goto LABEL_12;
        }
        totoal_files = bxx_num + 1;             // .b0x + .mdt file
[...]
\end{lstlisting}

After reverse engineering, we can determine how the trusted application is loaded from the normal world and create a script to merge the scattered files into a complete executable file. However, this executable file is not a directly runnable program but rather a shared object file.

\subsection{Emulator Design}

\begin{figure}[tbp]
  \centering
  \includegraphics[width=\columnwidth]{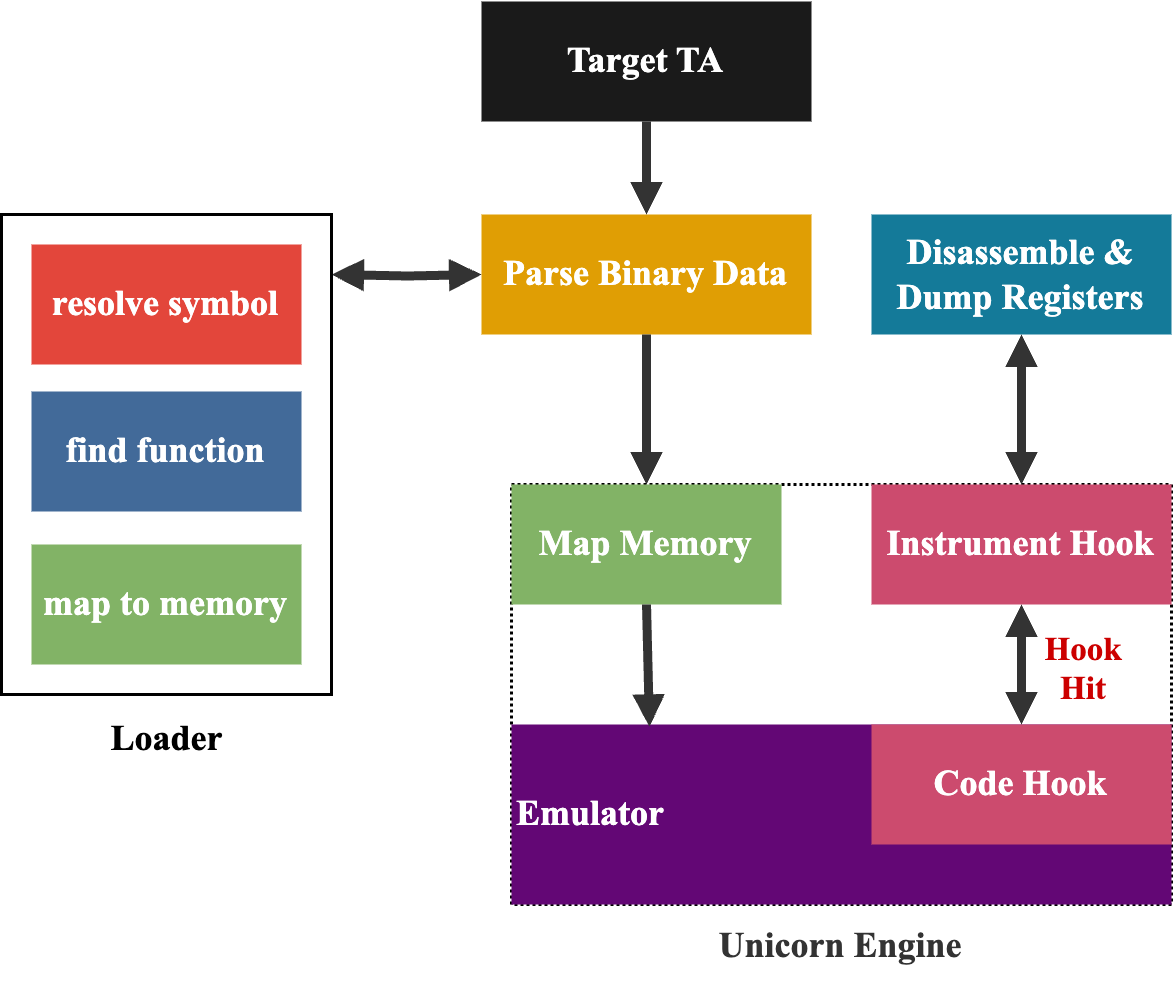}
  \caption{Emulation Architecture}
  \label{fig:unicorn-emu-loader}
\end{figure}

The emulator design utilizes the Unicorn engine. First, parse the binary file and map the memory area where the program will load into the emulator to ensure accurate address resolution during execution. After completing the mapping, instrument hooks to facilitate debugging and verify the correctness of the emulation architecture, as illustrated in Figure \ref{fig:unicorn-emu-loader}.

Select the one whose \texttt{p\_type} is \texttt{PT\_LOAD} in the program header structure in (Listing \ref{Elf64_phdr}), the emulator also needs to map to memory according to the program header. These memories include the code segment, bss segment, and data segment for normal executable programs. 

\begin{lstlisting}[language=C, caption=Elf64\_Phdr, label=Elf64_phdr]
    typedef struct {
       uint32_t   p_type;
       uint32_t   p_flags;
       Elf64_Off  p_offset;
       Elf64_Addr p_vaddr;
       Elf64_Addr p_paddr;
       uint64_t   p_filesz;
       uint64_t   p_memsz;
       uint64_t   p_align;
    } Elf64_Phdr;
\end{lstlisting}

Before starting the emulation, set the stack address and allocate sufficient space to ensure the program executes correctly according to the calling convention. With the memory area mapped, the emulator can now handle any function within the program's text segment. Identify the code’s starting and ending positions for emulation.

Additionally, if variables from other runtime programs need initialization or processing before function use, employ hooking technology. For instance, a hook can check if a parameter is \texttt{NULL} or verify a specific block of global memory for initialization. Before actual emulation, use hooks to prepare the environment, and dump output results when the emulator reaches a specific address. The emulator can also print the temporary register values during function calls to verify parameter correctness.

\begin{figure}[tbp]
  \centering
  \includegraphics[width=0.8\columnwidth]{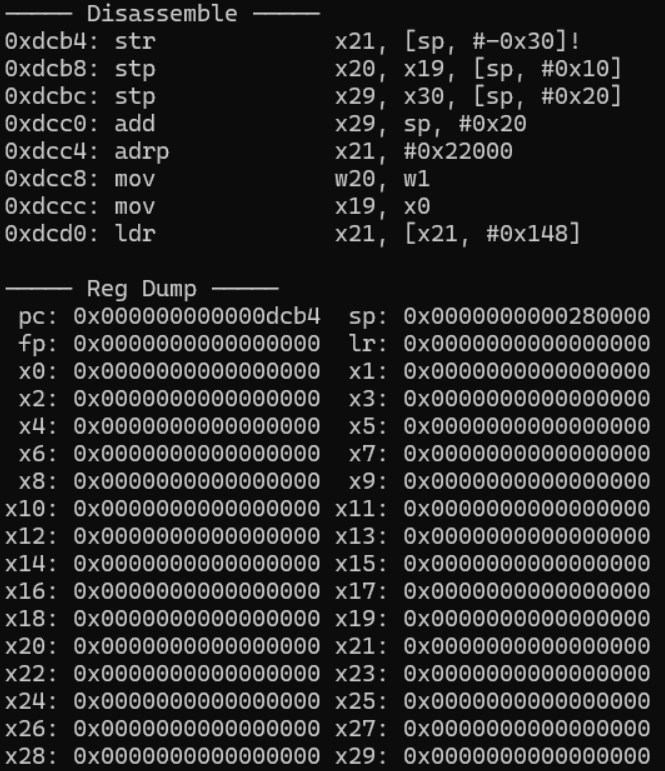}
  \caption{Instructions and Registers Dump}
  \label{fig:regdump}
\end{figure}

The hook content involves translating the executed machine code and confirming correctness through static analysis. Additionally, the emulator dumps the temporary register values during execution. This method helps verify whether the emulator executes instructions correctly, as illustrated in Figure \ref{fig:regdump}.

\subsection{Loader Design}

\begin{figure}[tbp]
  \centering
  \includegraphics[width=\columnwidth]{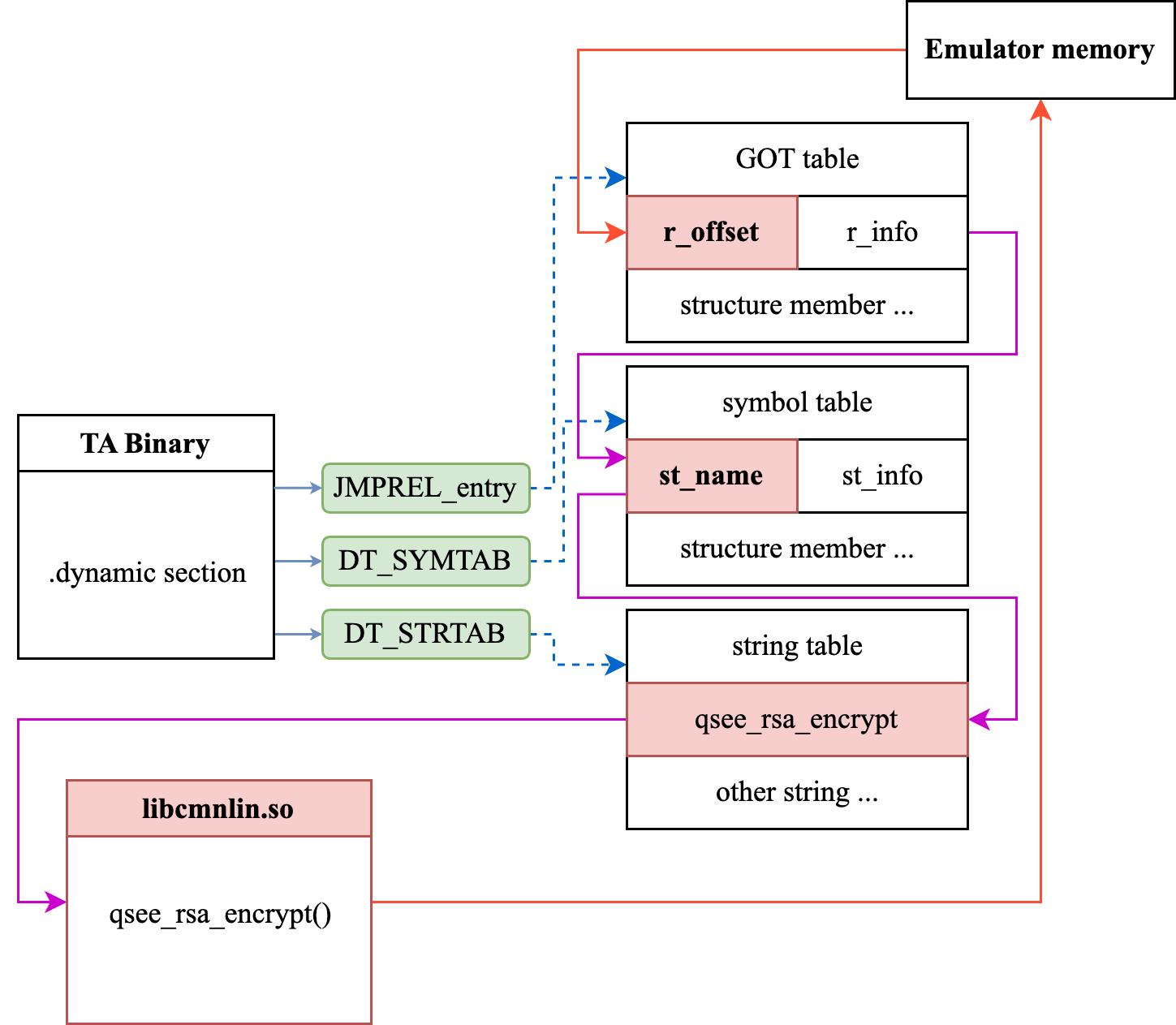}
  \caption{Loader Emulator}
  \label{fig:loader_design}
\end{figure}

The design of the loader, as shown in Figure \ref{fig:loader_design}, is necessary because the emulator must utilize the trusted application's dependency library within the system. This library, \texttt{libcmnlib.so}, includes various encryption and decryption functions such as \texttt{qsee\_rsa\_encrypt} and \texttt{qsee\_rsa\_decrypt}. To ensure the emulation environment closely mirrors the actual program execution, the functions and contents of this library must be integrated into the emulator. Since the trusted application is a dynamically linked binary, it will employ a lazy binding strategy to parse symbols and relocate functions, then update the corresponding entries in the GOT table through the linker. However, no directly usable linker is available on the system.

First, locate the .dynamic section in the trusted application and parse each entry to identify the address where \texttt{d\_tag} is \texttt{JMPREL}, which indicates the entry of the GOT table. The .dynamic entry with \texttt{d\_tag} as \texttt{DT\_SYMTAB} points to the symbol table, while the entry with \texttt{DT\_STRTAB} references the string table in the binary.

According to the documentation, under a 64-bit architecture, the relocation structure will be \texttt{Elf64\_Rela} as shown in Listing \ref{Elf64_Rela}. The member \texttt{r\_offset} specifies the address to be filled after resolving the symbol, while \texttt{r\_info} contains the index in the symbol table. To obtain the symbol table index, shift \texttt{r\_info} right by 32 bits, revealing its position in the symbol table.

\begin{lstlisting}[language=C, caption=Elf64\_Rela, label = Elf64_Rela ]
    typedef struct {
       Elf64_Addr r_offset;
       uint64_t   r_info;
       int64_t    r_addend;
    } Elf64_Rela;
\end{lstlisting}

Having obtained the location of the symbol table from the parsing of the \texttt{.dynamic} section entry, locate the corresponding symbol by adding the size of \texttt{Elf64\_Sym * index}. Retrieve the function's name by accessing the \texttt{st\_name} field in the structure shown in Listing \ref{Elf64_Sym}. The \texttt{st\_name} member provides the offset in the string table. The overall process for resolving function symbols is as follows:

1. Parse the ELF \texttt{.dynamic} section to identify the symbol table, string table, and GOT table.

2. Locate the symbol table entry based on the index from the GOT table.

3. Retrieve the function name from the string table using the \texttt{st\_name} offset in the symbol table.

\begin{lstlisting}[language=C, caption=Elf64\_Sym, label = Elf64_Sym ]
    typedef struct {
       uint32_t      st_name;
       unsigned char st_info;
       unsigned char st_other;
       uint16_t      st_shndx;
       Elf64_Addr    st_value;
       uint64_t      st_size;
    } Elf64_Sym;
\end{lstlisting}

\begin{figure}[tp]
  \centering
  \includegraphics[width=\columnwidth]{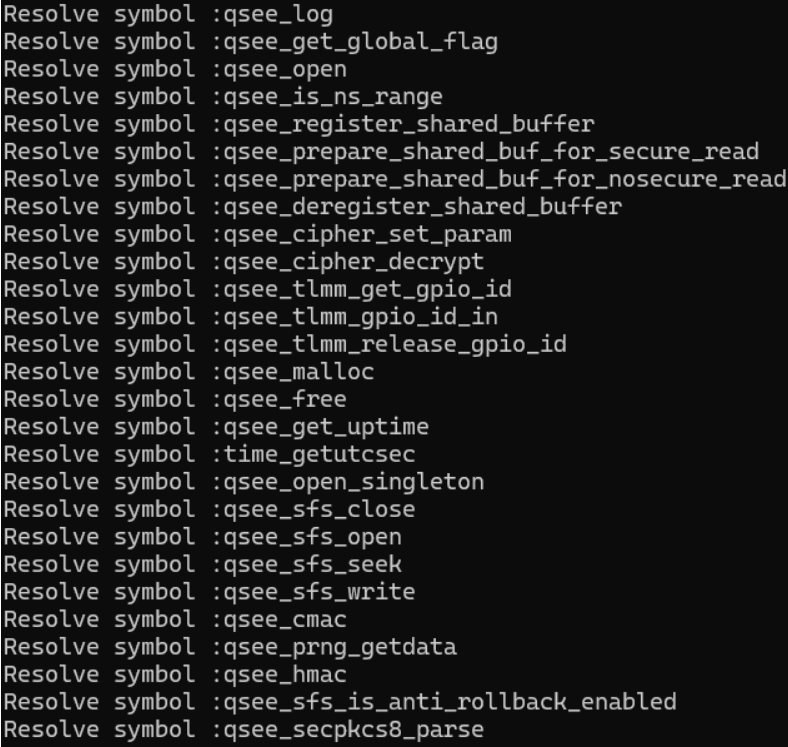}
  \caption{Symbol Resolving}
  \label{fig:resolve-symbol}
\end{figure}

Following the steps outlined, the function symbols are resolved sequentially. The results, shown in Figure \ref{fig:resolve-symbol}, reveal that these functions are all implemented in \texttt{libcmnlib.so}.

\subsection{Reverse Qualcomm Widevine}
Widevine's digital rights management (DRM) solution offers robust capabilities for licensing, protecting, and securely distributing content across various consumer devices. This enables consumers to enjoy high-quality digital content securely, whether on a smartphone, tablet, smart TV, or other devices. Widevine's solution is crucial for digital content owners, multi-service operators, and media providers, as it safeguards their content from unauthorized access and use, ensuring legal distribution. Regardless of the device, Widevine's technology guarantees seamless and secure delivery of purchased or subscribed content. By providing the protection technology, content providers can invest more in content creation and service enhancements, improving the consumer experience. According to the official description \cite{Widevine_offical}, over 5.0 billion devices currently use this technology, with major manufacturers like Google, Samsung, and Qualcomm incorporating it into their products.

Reverse engineering Qualcomm's trusted application alongside a Widevine trusted application highlights Widevine's crucial role in DRM. User data is primarily transmitted via commands, with a command handler in the trusted application managing these instructions. The main focus during reverse analysis is to locate the \texttt{cmd\_handler} and understand how to process the commands.

Begin the analysis by identifying the program entry. Since the trusted application becomes a shared object without a main entry point, it is utilized and called by the TEE through individual function units. Therefore, the reverse engineering process should start by locating the \texttt{CElF\_fileinvoke()} function in the binary file and then identifying the command handler in Figure \ref{cmd_handler}.

\begin{lstlisting}[language=C, caption= Command Handler, label = cmd_handler ]
_DWORD *__fastcall cmd_handler(_DWORD *result, unsigned int a2, __int64 a3, unsigned int a4)
{
  unsigned int v4; // w8

  if ( result && a3 )
  {
    v4 = *result & 0xFFFF0000;
    switch ( v4 )
    {
      case 0x60000u:
        return (_DWORD *)widevine_dash_cmd_handler(result, a2, a3, a4);
      case 0x50000u:
        return (_DWORD *)drmprov_cmd_handler();
      case 0u:
        return (_DWORD *)tzcommon_cmd_handler();
    }
  }
  return result;
}
\end{lstlisting}

\begin{figure}[tbp]
  \centering
  \includegraphics[width=0.6\columnwidth]{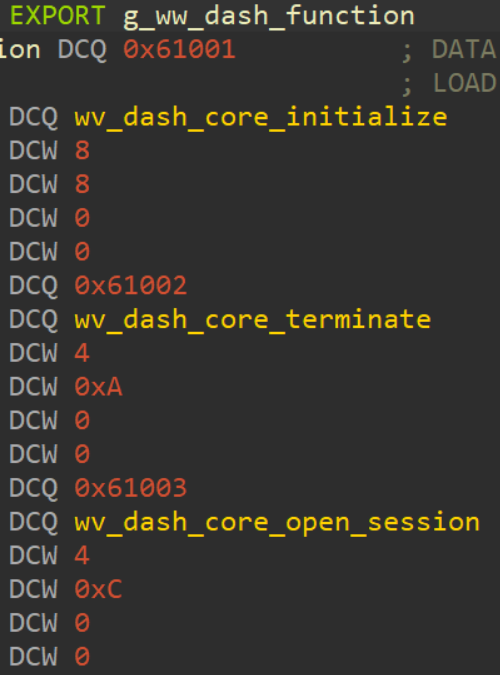}
  \caption{Widevine Command Table}
  \label{fig:cmd_table}
\end{figure}

When further in-depth reverse analysis of \texttt{widevine\_dash\_cmd\_handler()}, there is a pointer pointing to a table. After analyzing and inferring some types, it is a dash command table, which records command id, command pointer, etc, and other information in (Listing \ref{fig:cmd_table}). 

In a trusted application, the reliance on third-party function libraries can complicate function relocation and increase dependency on external libraries. To mitigate these issues, many functions are implemented directly within the trusted application or a single API library. For instance, during reverse analysis may discover that the function \texttt{memcpy()} is implemented internally rather than relying on \texttt{memcpy()} from glibc or other libraries.

A notable feature in \texttt{dash\_command\_table} is that the functions it contains serve as wrappers, with \texttt{OEMCrypto} as a prefix for the processed functions. For example, calling \texttt{wv\_dash\_core\_security\_patch\_level()} ultimately invokes \texttt{OEMCrypto\_Security\_Patch\_Level()} as shown in Figure \ref{OEM_patch_level}.

\begin{lstlisting}[language=C, caption= \texttt{wv\_dash\_core\_security\_patch\_level()} , label = OEM_patch_level ]
__int64 __fastcall wv_dash_core_security_patch_level(char *a1, __int64 a2)
{
  __int64 result; // x0
  char v5; // w13
  char v6; // w14

  result = OEMCrypto_Security_Patch_Level(a2 + 4); // -> Function implement place
  
  [....]
  
  return result;
}
\end{lstlisting}

\section{Evaluation Result Analysis}

This chapter introduces the experiment of the proposed scheme, connecting the emulator with the fuzzer to discover existing or new vulnerabilities. It details how to integrate the fuzzer with the proposed method and how to select mutation targets as user inputs for different code fragments.

\subsection{Environment}

Table \ref{setup} outlines the experimental environment, which includes the Ubuntu 22.04 operating system, an i7-13700K CPU, and 64GB of RAM. The fuzzer is AFL++ \cite{AFL++} in unicorn mode. AFL++ is an optimized variant of the AFL \cite{AFL} fuzzer, offering various modes for fuzzing, including qemu mode, unicorn mode, and frida mode.

\begin{table}[h]
    \caption{Environment Setup}
    \centering
    \begin{tabular}{|l|l|}
    \hline
    CPU & Intel i7-13700K  \\ \hline
    Memory & 64GB\\ \hline
    OS & Ubuntu 22.04 LTS\\ \hline
    Fuzzer & AFL++\\ \hline
    \end{tabular}
    \label{setup}
\end{table}

\subsection{American Fuzzy Lop plus plus (AFL++)}

Fuzzing testing aims to uncover security vulnerabilities in software by supplying unexpected or invalid inputs and analyzing the outcomes. AFL++ employs various techniques to generate new test cases and explore the target program's input space. Prominent methods include bit flipping, dictionary-based mutations, and havoc. These techniques are detailed below:

\begin{itemize}
    \item Bitflip: Bit flipping is a fundamental mutation technique used in fuzzing. This method involves creating new test cases by flipping individual bits in the input data. Bit flipping helps uncover bugs related to specific bit patterns and can trigger edge cases in the target program.

    \item Dictionary: Dictionary-based mutations utilize predefined words or byte sequences, often derived from input format or protocol specifications, to generate new test cases. This approach leverages knowledge about the input structure to create meaningful and potentially more effective test cases.
    
    \item Havoc: Havoc represents a more aggressive and random mutation strategy. It applies a broad range of mutations to the input data in a less systematic but more extensive manner. The goal of havoc is to produce inputs that cause the target program to behave unpredictably, potentially revealing hidden bugs.
\end{itemize}

Based on the vulnerability \cite{CVE-2021-0592} discovered by a previous researcher in widevine \cite{Breaking_Widevine}, we will apply our proposed scheme to emulate and conduct fuzzing testing to identify this vulnerability. The issue exists in \texttt{decrypt\_CTR\_unified()}. This vulnerability arises because the user can control the buffer address without any boundary checks, leading to a potential out-of-bounds write vulnerability.

For fuzzing testing, identifying the input positions within the function and verifying whether the input is user-controllable are crucial. Take \texttt{decrypt\_CTR\_unified()} as an example. The function performs numerous parameter checks at the beginning. By using cross-references to trace back to the function's callers, we can determine which parameters are user-controllable, as shown in (Figure \ref{decrypt_CTR_caller}). Parameters such as \texttt{do\_decrypt}, \texttt{enc\_buf + offset}, and \texttt{out\_buf + offset} are passed by the user.

\begin{lstlisting}[language=C,caption= Caller of \texttt{decrypt\_CTR\_unified()}, label = decrypt_CTR_caller]
      ret = decrypt_CTR_unified(
              session_id,
              (char *)(enc_buf + offset),
              data_len_to_dec,
              do_decrypt,
              v55,
              v68,
              (char *)(out_buf + offset),
              out_len,
              a7,
              buf_meta,
              max_length,
              1);
    }
\end{lstlisting}


When entering \texttt{decrypt\_CTR\_unified()}, there are many checks at the starting point of the function. Most of the checks are to determine whether its content is empty. If this first check fails, it will be interrupted and the program will not be executed further in (Listing \ref{decrypt_CTR_unified()}). Therefore, the analysis of which parameters should be filled correctly is necessary to be checked. Before starting to fuzz the target, the hook instrumentation is required here for processing. For example, session\_id can be set to any value less than 0x32 as it does not affect the program code flow and subsequent use and has no related functions.

\begin{lstlisting}[language=C,caption=\texttt{decrypt\_CTR\_unified()}, label = decrypt_CTR_unified()]
  if ( session_id > 0x32
    || (v19 = session_table[2 * session_id]) == 0
    || !data_len_to_dec
    || a6 > 0xF
    || !buf_meta
    || !a9
    || !outsample
    || !a5
    || !insample
    || !a12 )
  {
    qsee_log(8LL, "Error: decrypt_CTR_unified: NULL pointers! p_session_ctx = 0x%x", (unsigned int)v19);
    qsee_log(
      8LL,
      "Error: subsample_in_data_addr = 0x%x, subsample_length = 0x%x",
      (unsigned int)insample,
      data_len_to_dec);
    qsee_log(8LL, "Error: iv = %x, out_buffer = 0x%x", (unsigned int)a5, (unsigned int)buf_meta);
    qsee_log(8LL, "Error: block_offset %d is larger than 16! ", a6);
    goto LABEL_14; // Error abort !
  }
\end{lstlisting}

User-controllable parameters will become objects that need to be mutated during fuzzing in (Listing \ref{hook_decrypt_CRT}). As for parameters that are not controllable by the user, the hooking technology mentioned earlier can allow the program to execute the function normally.

\begin{lstlisting}[language=C,caption=\texttt{decrypt\_CTR\_unified()} Setup, label = hook_decrypt_CRT]
void setup_decrypt_CTR_unified(uc_engine *uc){

	srand(time(NULL));
	int offset; 
    //user control offset , final_buf = buf + offset 
	offset = rand() % 0x1200;

	unsigned int x0 = 0;
	unsigned long insample = INSAMPLE + offset ;
	unsigned int x2 = 0x100;
	unsigned int x4 = 0xff;
	unsigned int x5 = 0;
	unsigned long outsample = OUTSAMPLE + offset;

	void *x7 = 0x500000 ;
	uc_reg_write(uc,UC_ARM64_REG_X0 , &x0);
    // 0x341E8 = section_table
	mem_write_wrapper(uc , 0x341E8 , 1);
	uc_reg_write(uc,UC_ARM64_REG_X1 , &insample);
	uc_reg_write(uc,UC_ARM64_REG_X2 , &x2);
	uc_reg_write(uc,UC_ARM64_REG_X4 , &x4);
	uc_reg_write(uc,UC_ARM64_REG_X5 , &x5);
	uc_reg_write(uc,UC_ARM64_REG_X6 , &outsample);
	uc_reg_write(uc,UC_ARM64_REG_X7 , &x7);
 
    // parameter on stack , (aarch64 calling convention)
	
	mem_write_wrapper(uc , 0x280008 , 0xf1);
	mem_write_wrapper(uc , 0x280000 , 0xf0);
	mem_write_wrapper(uc , 0x280010 , 0x1f2	);
	mem_write_wrapper(uc , 0x280018 , 0xf3	);
 }
\end{lstlisting}

\subsection{Fuzzing Harness}

After the emulation step is completed and confirmed that the function can be executed normally, the next step is to connect the target function with the fuzzer. Harness is the interface between the fuzzer and the target function. The fuzzer needs to know which memory space of the emulator to put the mutated input. For example, if the function parameter receives a pointer type, the memory area pointed to by the pointer needs to take place as the input address. The fuzzer in this research is set to be executed in a persistent mode in (Figure \ref{persistent_fuzz}), which means that the state of each fuzz remains the same as before fuzzing, only the input of the mutation changes.

\begin{lstlisting}[language=C,caption=Prepare for Persistent Fuzzing, label = persistent_fuzz]
[...]
    setup_decrypt_CTR_unified(uc); //init args
    // For persistent mode, need to set up stack and memory each time.
    uc_reg_write(uc, UC_ARM64_REG_PC, &CODE_ADDRESS); // Set the instruction pointer back
    // Need a valid c string, make sure it never goes out of bounds.
    input[input_len-1] = '\0';
    // Write the testcase to unicorn.
    uc_mem_write(uc, INPUT_LOCATION , input, input_len);
    // store input_len
    current_input_len = input_len;
[...]
\end{lstlisting}
\subsection{Crash Analysis}

Since the code does not adopt any checking mechanism on the memory boundary, and if the \texttt{do\_decrypt} parameter is 0, it will directly enter the memory copy function, and \texttt{memcpy()} simply copies the contents of the source address to the contents of the destination address. The copy went to a place in memory that was not mapped, causing a crash.

This research confirmed the correctness of the emulation in terms of emulation and successfully connected to the fuzzer for fuzzing testing. Before performing fuzzing testing, the stability of the emulator was thoroughly verified. This was done by using the debug feature to ensure that the function returns a reasonable return value before executing the fuzzing target. Being able to effectively discover known vulnerabilities as a verification method can help researchers try to discover vulnerabilities in TAs in a partially emulated method.

\subsection{Comparision with Relate Works}

Table \ref{compare} presents a comparative analysis of different approaches focused on fuzzing and emulating TAs within TEEs. The comparison highlights the scheme used, the fuzzer employed, how each approach handles dependency issues, and the evaluation outcomes. Below is a detailed explanation based on the table's content:

\begin{table*}[ht!]
    \caption{Comparison with Related Works }
    \label{compare}

    \centering
    
\begin{tabularx}{\textwidth}{|c|X|c|X|X|}
    \hline
     & Scheme & Fuzzer & Dependency issue & Evaluation  \\
    \hline
    TEEzz\cite{10179302} & Physical device fuzzing & Based on AFL++  & None & One CVE in OP-TEE\\
    \hline
    PARTEMU\cite{247658} & Full system emulation & AFL & Full emulation system to solve this issue & Crashes in TAs and TEEOS \\
    \hline
    DTA\cite{song2024dta} & Emulation with Qemu & AFL++ & Patching TA's assembly & Crashes in a sample vulnerable TA\\
    \hline
    SimTA\cite{busch2020finding}  & Argr-based emuation & None & Prototype method to archive parameters & N-Day rediscover and 1-day rediscover \\

    \hline
    Ours & Emulation with Unicorn & AFL++ & Loader design for symbol resolving & N-Day rediscover \\
    \hline
\end{tabularx}
    
\end{table*}

\noindent\textbf{TEEzz} employs \textit{physical device fuzzing}, utilizing a fuzzer based on AFL++. This approach is advantageous because it operates directly within the physical environment, ensuring that all dependencies are naturally resolved without additional intervention. However, this reliance on physical devices introduces limitations in scalability and complexity in setup, making it less flexible than emulation-based approaches. Despite these challenges, TEEzz successfully identified a known Common Vulnerability and Exposure (CVE) within OP-TEE, demonstrating its practical efficacy.

\noindent\textbf{PARTEMU} adopts a \textit{full system emulation} approach, using AFL as its fuzzer. By emulating the entire ARM TrustZone environment, including both the Secure and Normal Worlds, PARTEMU provides a comprehensive platform for testing. This method effectively addresses dependency issues through full emulation, ensuring accurate replication of the real-world environment. However, this approach is resource-intensive and complex to configure, which may hinder its adaptability and ease of use. Additionally, PARTEMU's reliance on closed-source methods limits the reproducibility and flexibility of the system, posing challenges for wider adoption in the research community. Nevertheless, PARTEMU successfully identified crashes in TAs and the TEE Operating System (TEEOS), indicating its effectiveness in detecting critical vulnerabilities.

\noindent\textbf{DTA} utilizes \textit{emulation with QEMU}, coupled with AFL++ as the fuzzer. This approach relocates TAs from the Secure World to the Normal World, enabling easier fuzzing and testing without compromising the secure environment. To resolve dependency issues, DTA employs patching of the TA's assembly code, allowing it to function correctly in the emulated environment. However, the necessity of assembly patching adds complexity to the process and may limit the emulation's accuracy regarding the Secure World's unique conditions. Despite these challenges, DTA successfully identified crashes in a sample vulnerable TA, validating its approach.

\noindent\textbf{SimTA} is based on \textit{angr-based emulation}, focusing on selective symbolic execution rather than traditional fuzzing. This method allows for targeted analysis of specific execution paths within the TA environment. SimTA addresses dependency issues through a prototype method that archives parameters, simplifying the testing process. While SimTA successfully rediscovered N-Day vulnerabilities and identified a 1-day vulnerability, the lack of a traditional fuzzer limits its ability to explore execution paths comprehensively. Additionally, the reliance on symbolic execution can be time-consuming and computationally intensive, potentially reducing the efficiency of vulnerability discovery.

\noindent\textbf{Our approach} leverages \textit{emulation with Unicorn}, a lightweight and flexible CPU emulator, connect AFL++ as the fuzzer. This method balances the accuracy of physical device testing with the flexibility of emulation, offering a practical solution for efficiently testing TA functions. A key strength of our approach is the custom loader design for symbol resolving, which effectively handles dependencies within the emulated environment, ensuring a realistic and robust testing context. Unlike PARTEMU, our method does not require the full emulation of the entire system, making it less resource-intensive and easier to configure. Additionally, our approach avoids the complexities associated with assembly patching in DTA, providing a more straightforward and accurate testing process. Through this method, we successfully rediscovered known N-Day vulnerabilities, demonstrating the efficacy and reliability of our approach.

In summary, our method offers a significant advantage by providing an open-source, replicable, and efficient emulation and fuzzing testing strategy. It addresses the limitations observed in other approaches, such as the resource intensity of PARTEMU, the complexity of DTA, and the limited exploration capabilities of SimTA. Our approach represents a balanced and practical solution for emulating and testing trusted applications, contributing to the advancement of TEE security research.

\section{Conclusions and Future Works}
\label{ch:conclusion}

TEEs are integral to a wide range of modern products. However, the complexity and inherent limitations of these environments pose significant security challenges to the internal programs. This research extracted the trusted application from an actual Pixel 4 smartphone, fully analyzed it, and designed an emulation environment capable of executing arbitrary code fragments.

Emulating cross-architecture programs presents numerous challenges, including understanding program execution, addressing issues within the emulation environment, and resolving differences between the emulated and actual environments. For black-box target programs, data collection is only possible through reverse engineering, which involves analyzing assembly execution, memory mapping, and caller-callee relationships. TAs often contain complex logic, making it difficult to quickly assess vulnerabilities through manual analysis. Fuzzing testing mitigates this by reducing the manual effort required for vulnerability discovery.

The primary contribution of this research is the development of an emulation-based approach for fuzzing testing TAs in TEE. A debugger was integrated into the emulation environment to ensure consistency between the emulated and actual environments. A loader was also designed to map the functions used in the trusted application into the emulator. This research successfully demonstrated that arbitrary code fragments within the trusted application could be emulated and that fuzzing testing can effectively identify existing vulnerabilities.

The TAs analyzed in this research were extracted from physical mobile devices. Future work will focus on automating the extraction and analysis of firmware to locate TAs. As different firmware versions implement varying functions, further research will explore methods for adapting the emulator to handle these differences. Additionally, developing techniques to reduce the cost of reverse engineering and expedite the completion of harnesses before fuzzing testing will be a key area of study.

\ifCLASSOPTIONcaptionsoff
  \newpage
\fi

\bibliographystyle{IEEEtran}
\bibliography{citation}


\vfill

\end{document}